\documentclass[useAMS,usenatbib,usegraphicx]{mn2e}
\usepackage{times}

\def\rxte{{\sl RXTE}}

\def\vlt{{\em VLT}}
\def\ultracam{{\sc ultracam}}

\def\ltsim{\mathrel{\hbox{\rlap{\hbox{\lower4pt\hbox{$\sim$}}}\hbox{$<$}}}}
\def\gtsim{\mathrel{\hbox{\rlap{\hbox{\lower4pt\hbox{$\sim$}}}\hbox{$>$}}}}

\def\Msun{M$_{\odot}$}

\def\gravrad{$R_{\rm G}$}

\def\gx339{GX~339--4}
\def\swiftj1753{SWIFT~J1753.5--0129}
\def\xtej1118{XTE~J1118+480}
\def\av{$A_{\rm V}$}
\def\rin{$r_{\rm in}$}

\title[Rapid optical/X-ray correlations in GX~339--4]{Rapid optical and X-ray timing observations of GX~339--4: flux correlations at the onset of a low/hard state\thanks{Based on observations carried out in ESO programmes 079.D-0535 and 279.D-5021, and \rxte\ Proposal number 93119.}}

\author[P. Gandhi et al.]{P. Gandhi,$^1$ 
K. Makishima,$^{1,2}$ M. Durant,$^3$ A.C. Fabian,$^4$ V.S. Dhillon,$^5$ T.R. Marsh,$^6$ \newauthor J.M. Miller,$^7$ T. Shahbaz$^3$ and H.C. Spruit$^8$\\
$^{1}$RIKEN Cosmic Radiation Lab, 2-1 Hirosawa, Wakoshi, Saitama 351-0198, Japan\\
$^{2}$Department of Physics, University of Tokyo, 7-3-1 Hongo, Bunkyo-ku, Tokyo 113-0033, Japan\\
$^{3}$Instituto de Astrof\'{i}sica de Canarias, La Laguna, E38205 Tenerife, Spain\\
$^{4}$Institute of Astronomy, Madingley Road, Cambridge, CB3 0HA\\
$^{5}$Department of Physics and Astronomy, University of Sheffield, Sheffield S3 7RH\\
$^{6}$Department of Physics, University of Warwick, Gibbet Hill Road, Coventry, CV4 7AL\\
$^{7}$Department of Astronomy, University of Michigan, 500 Church Street, Ann Arbor, MI 48109, USA\\
$^{8}$Max-Planck-Institut f\"{u}r Astrophysik, Postfach 1317, 85741 Garching bei M\"{u}nchen, Germany
}

\begin{document}

\date{
Accepted 2008 July 8.  Received 2008 July 8; in original form 2008 June 21
}

\pagerange{\pageref{firstpage}--\pageref{lastpage}} 
\pubyear{2008}

\maketitle
\label{firstpage}

\begin{abstract}
We present the discovery of optical/X-ray flux correlations on rapid timescales in the low/hard state of the Galactic black hole GX 339--4. The source had recently emerged from outburst and was associated with a relatively-faint counterpart with mag $V$$\approx$17. The optical (\vlt/ULTRACAM) and X-ray (\rxte/PCA) data show a clear positive cross-correlation function (CCF) signal, with the optical peak lagging X-rays by $\sim$ 150 ms, preceded by a shallow rise and followed by a steep decline along with broad anti-correlation dips. Examination of the light curves shows that the main CCF features are reproduced in superpositions of flares and dips. The CCF peak is narrow and the X-ray auto-correlation function (ACF) is broader than the optical ACF, arguing against reprocessing as the origin for the rapid optical emission. X-ray flaring is associated with spectral hardening, but no corresponding changes are detected around optical peaks and dips. The variability may be explained in the context of synchrotron emission with interaction between a jet and a corona. The complex CCF structure in GX~339--4 has similarities to that of another remarkable X-ray binary XTE~J1118+480, in spite of showing a weaker maximum strength. Such simultaneous multi-wavelength, rapid timing studies provide key constraints for modeling the inner regions of accreting stellar sources.
\end{abstract}

\section{Introduction}

\gx339 is an X-ray binary system hosting one of the most promising Galactic black hole (BH) candidates, with a mass function of $\sim 6$~\Msun\ \citep{hynes03_gx339}. It is classified as a micro-quasar, based on tight radio/X-ray correlations found over a large dynamic range in source luminosity \citep[e.g. ][]{gallo03}. Extensive X-ray studies have revealed a rich timing structure \citep[e.g. ][]{dunn08, belloni05, nowak99, miyamoto91}. In its low state, the source is typically associated with an optically-bright counterpart, during which flickering on timescales as short as 10~ms has been observed. Rare transitions to the X-ray--off state have enabled placing some constraints on the much fainter optical companion star \citep{shahbaz01}. And during X-ray outbursts, the optical and X-ray fluxes display complex correlations as well as anti-correlations over extended timescales \citep[e.g. ][]{makishima86,russell06}.

Rapid timing observations simultaneous over a broad energy range remain as one of the lesser-explored aspects of \gx339. The last published multi-wavelength studies to probe sub-second timescales were carried out over 20 years ago (but see Spruit et al. in prep.). Besides high state observations by \citeauthor{makishima86}, \citet[][]{motch82, motch83, motch85} studied the source during a low- to high-state transition, and found an anti-correlated cross-correlation function (CCF) signal with optical leading X-rays by a few seconds. Although there was an indication in the low-frequency light curves that the optical and 13--20 keV X-ray fluxes were correlated at a significance of about 98 per cent, no obvious high frequency ($\ltsim 20$ s) correlation was uncovered at any of the selected energies. 

The Galactic-halo X-ray binary \xtej1118\ has recently been a target of extensive, simultaneous multi-wavelength coverage. Among its interesting properties, a complex optical/X-ray CCF has been found \citep[e.g. ][]{kanbach01, spruitkanbach02}, including positive and negative correlation components and an optical-vs.-X-ray peak lag of $\sim 0.5$ s. Complex CCFs are now being found in other sources as well (\citealt{durant08}). These provide important time domain constraints for physical models.

In this Letter, we present the first simultaneous optical/X-ray timing analysis of \gx339\ on rapid timescales of $\sim 50-130$ ms in a low/hard state associated with a relatively-faint optical counterpart. A clear cross-correlation signal is detected with the optical peak lagging X-rays by $\sim 150$ ms. The CCF has a complex pattern with some similarities as well as differences to the CCF found for \xtej1118, and is likely to be result of distinct interacting accretion/ejection components. Full details of the timing and spectral analysis are presented in a forthcoming paper.

\section{Observations}

The triple-beam optical camera \ultracam\ \citep{ultracam}, capable of high-speed photometry at up to 500 Hz, was mounted on the {\sl Very Large Telescope (VLT)} as a visitor instrument during Jun 2007. We carried out three 1~h long observations of \gx339 on alternate nights of UT Jun 14th, 16th and 18th (hereafter referred to as Nights 3, 2 and 1 respectively in order of improving weather), simultaneously with the {\sl Rossi X-ray Timing Explorer (RXTE)} satellite. Only the final night [Night 1] was photometric, while the other two had variable transparency, Night 3 being the worst. This period fell a few weeks after the source had returned to the low/hard state following a large X-ray outburst \citep{kalemci07} and also coincided with rising optical emission \citep{buxtonbailyn07}. Our choice of time resolution was governed by the need to obtain a good signal:noise under prevailing atmospheric conditions. The final values used were $\approx 50, 133$ and 136 ms on Nights 1, 2 and 3 respectively. Data calibration and relative photometry (with respect to a brighter comparison star observed simultaneously) was carried out with the \ultracam\ pipeline v. 8.1.1. Three-filter simultaneous observations are possible and we used $u'$, $g'$ and $r'$, but much longer exposures were required in $u'$ to obtain comparable signal:noise; consequently, the $u'$ data is not considered further in this Letter. Optical spectro-photometry carried out with the \vlt/FORS2 instrument on Night 1 gives $F_{\lambda}^{5000 \AA}= 6\times 10^{-16}$ erg s$^{-1}$ cm$^{-2}$ \AA$^{-1}$ $(V_{\rm Vega}\approx 17)$. Correction for Galactic extinction of \av$\approx$3.3 implies $\lambda L_\lambda^{5000 \AA}\approx 4.6\times 10^{35} (d/8\ {\rm kpc})^2$ erg s,$^{-1}$ \citep[distance from ][]{zdziarski04}.

\rxte\ observed the target in its canonical {\tt GoodXenon} and {\tt Standard} PCA modes, and in a 32-s on-off rocking mode with HEXTE cluster 1. Recommended HEADAS v. 6.4 procedures were followed for data reduction and extraction of light-curves and spectra, including the latest calibration information and background model corrections. A simple hard power-law with photon-index $\Gamma=1.65\pm0.02$ provided a statistically acceptable fit to the spectra for energies of 3--200 keV. The source had a flux $F_{2-10}=1.6\times 10^{-10}$ erg s$^{-1}$ cm$^{-2}$, $=> L_{2-10}=1.2\times 10^{36}$ erg s$^{-1}$, implying an optical($V$):X-ray($2-10$ keV) luminosity ratio of about 40 per cent. The low flux and hard power-law are characteristic of the source in the low/hard state. \citet{tomsick08} find that the X-ray flux reached a minimum during our observation period.

The source showed a high fractional variability amplitude in X-rays, approaching 50 per cent in the full-band PCA energy range (above the value expected from Poisson fluctuations) on all nights. In the optical, the net rms variability over the full light curves was $\approx 13$ and 15 per cent in the $g'$ and $r'$ filters, respectively.

\section{Results : the cross-correlation function}

\begin{figure*}
  \begin{center}
    \includegraphics[angle=90,width=15cm]{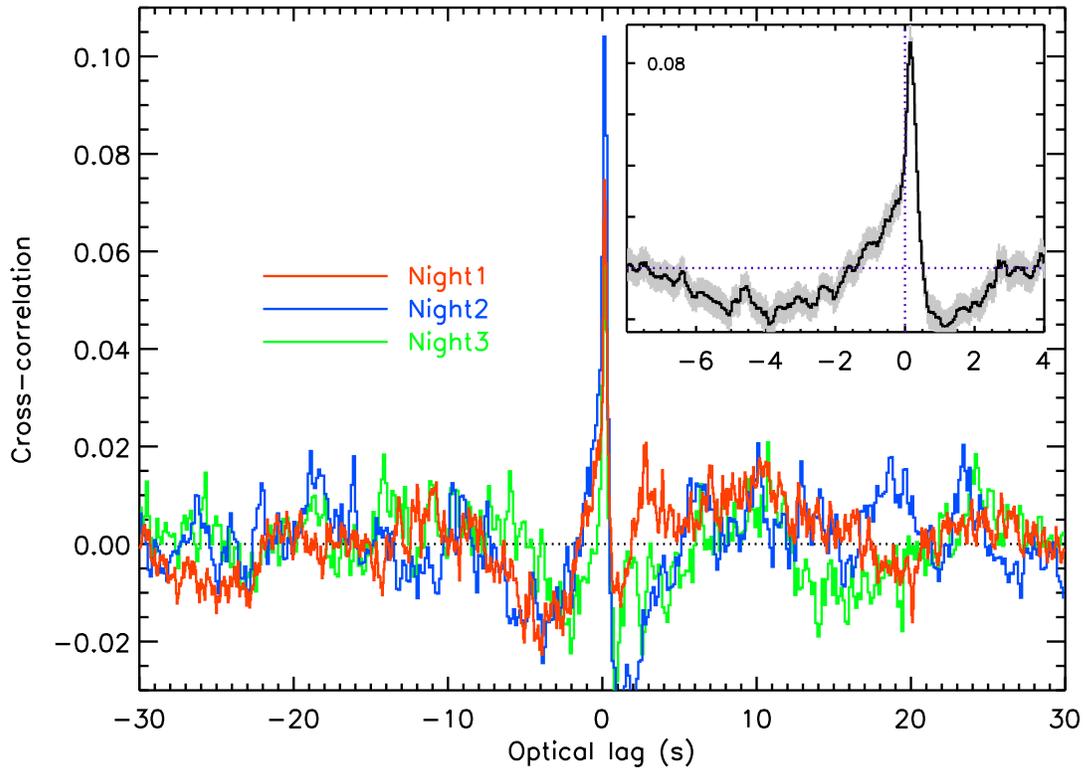}
\caption{The $r'$ vs. X-ray full-band PCA cross-correlation for data on all three nights, obtained from simultaneous light curve sections of 60 s length. A positive delay (in this case peaked at $\approx 150$ ms) implies that optical lags X-rays. The inset shows a zoom-in average CCF of Nights 1 and 2 interpolated onto the fastest timescale of 50 ms in order to clearly illustrate the delay. The shaded region is the average scatter computed in an ensemble of light curve sections.
 \label{fig:crosscorr}}
  \end{center}
\end{figure*}

The net optical and X-ray light curves were translated to a common Barycentric frame, and cross-correlated on the fastest optical time resolutions available on each night. The absolute and relative timing accuracies of \ultracam\ are $\sim 1$ ms and 50 $\mu$s, respectively \citep{ultracam}, much better than the smallest timescales in our light curves. The main result of our work is shown in Fig.~\ref{fig:crosscorr}. The CCF shows a single, significant peak at an optical lag of $\sim$ 150 ms. The peak itself has a narrow core, with a shallow rise from $\sim -1.5$ s to 0 s, and a steep decline from 150 ms to $\sim 0.5$ s. Weaker, but significant anti-correlation troughs are centred at $\sim -4$ s and 1 s. Each of these structures is visible in all the observations (in spite of some clear inter-night variation), suggesting that each of them is real. The peak narrowness and position are constant between the three nights, within the $\sim 50-130$ ms resolution available. The $g'$ data shows a very similar CCF to the $r'$/X-ray one presented. 

\begin{figure*}
  \begin{center}
    \includegraphics[angle=90,width=6.5cm]{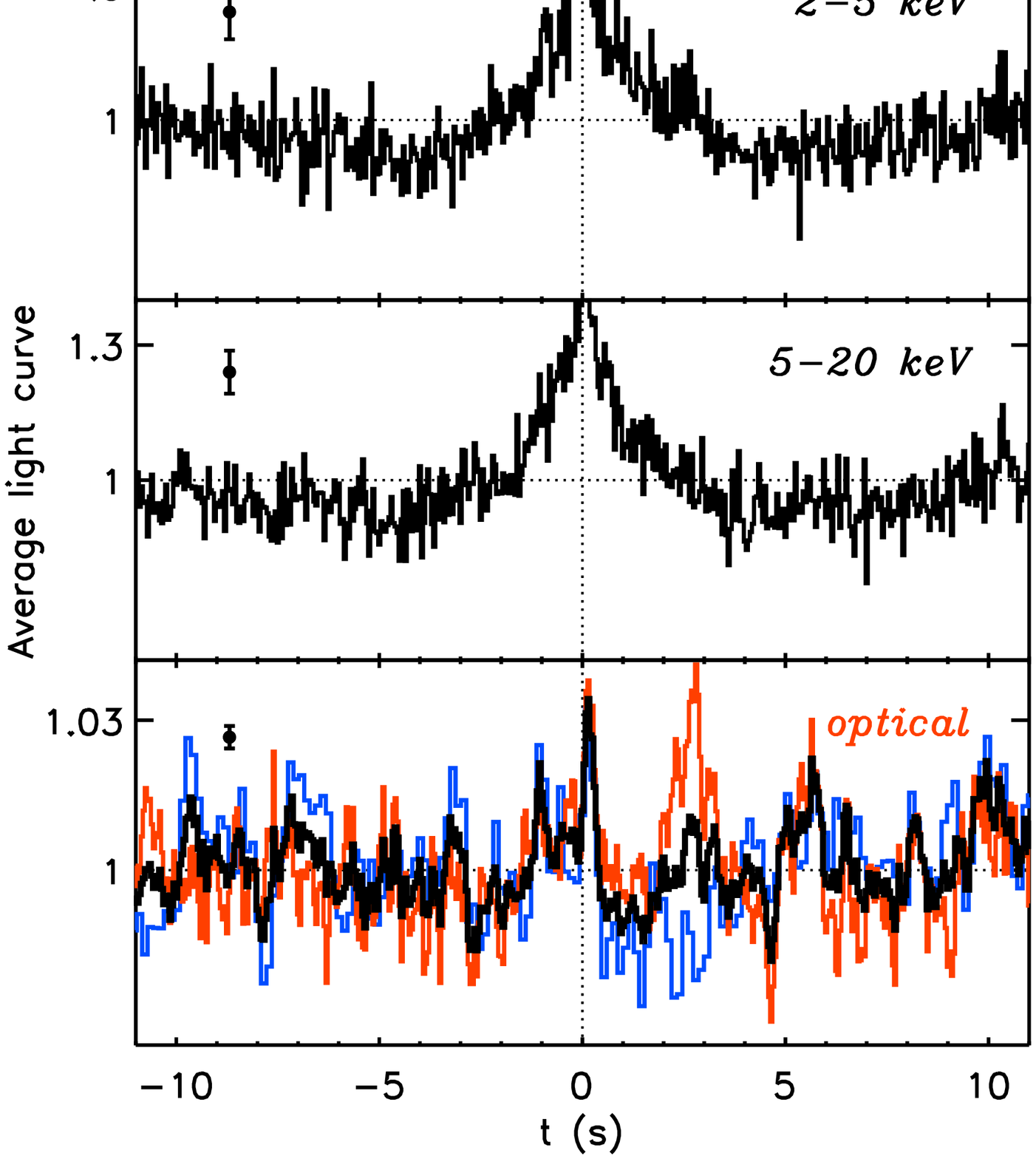}
    \includegraphics[angle=90,width=6.5cm]{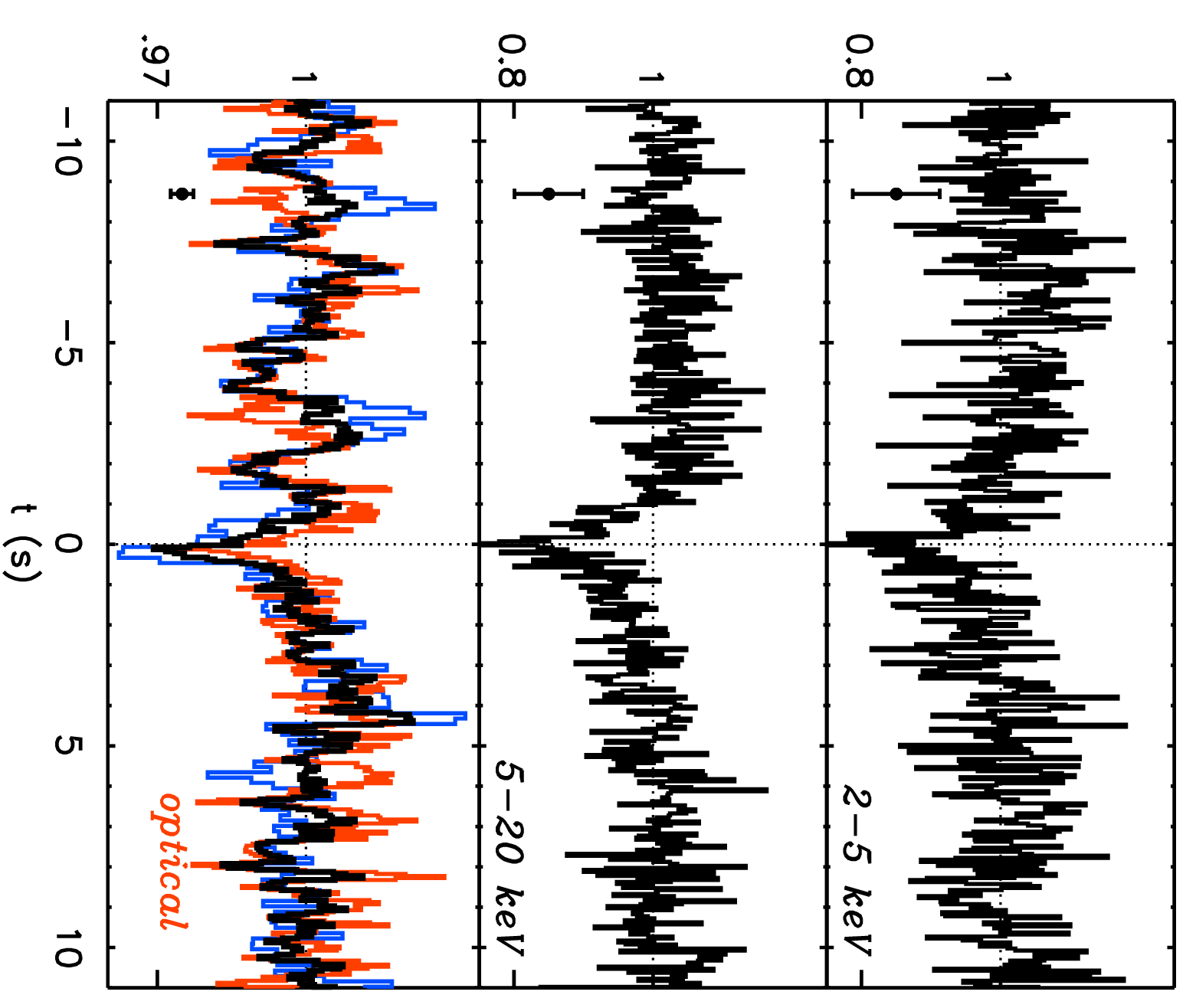}
\caption{Averaged 2--5 keV, 5--20 keV and optical ($r'$) light curves around full-band PCA X-ray flares {\bf\em (left)} and dips {\bf\em (right)}. Optical light curves from Nights 1 and 2 are shown in red and blue; averages from both nights are shown in black. For making this plot, full-band extrema are selected according to the method of \citet[][ cf. their Fig.~9]{malzac03}. A flare (or dip) must be at least $f$=2 times above (or $1/f$ times below) the local X-ray mean in a running $t_m$=32 s long section, and is also required to be the local extremum within a contiguous segment of $\pm t_p$=8 s (this effectively selects significant flares only). Corresponding light curves sections in other bands are then normalized to their local means before being averaged. Error bars show typical Poisson uncertainties.
 \label{fig:likemalzac}}
  \end{center}
\end{figure*}

The asymmetric shallow rise and steep decline of the CCF is clearly reminiscent of that seen in \xtej1118\ \citep{kanbach01}, but \lq mirror-imaged\rq\ about a vertical axis and shifted to a lag of 0.15 s. This lag was confirmed by constructing averaged optical light curves around micro- (local) flares and dips selected in the full-band X-ray data. Fig.~\ref{fig:likemalzac} shows the resultant optical (as well as 2--5 and 5--20 keV X-ray) superpositions around several hundred X-ray peaks and dips. An optical extremum appears at $\sim 150$ ms lag on all nights (though only the best weather Nights 1 and 2 are shown). Furthermore, the peaks light curve clearly shows lower-than-average troughs, as well as a preceding rise of the optical, all matching the CCF within $t$=$\pm$2 s. There are also some matches beyond this range, including the peaks local maximum at $t$=$+10$ s. But strong intrinsic (not Poisson) variability dominates, especially in the higher-resolution Night 1 data. This suggests that more flaring is present on timescales smaller than those probed by us. 

There may be several reasons why the previous observations of \citet{motch83} and \citet{makishima86} did not find a positive CCF signal. Firstly, the CCF strength itself probably evolves between the different states probed: a high-state during the observation of \citeauthor{makishima86} and an optically-bright [$V\approx 15.4$] low state in \citeauthor{motch83} During our observations, the CCF in Fig.~\ref{fig:crosscorr} has a peak strength of only $\sim 0.1$; a perfect correlation would show a peak of 1 (for comparison, \xtej1118\ has a peak of 0.4). It is also possible that the length of the simultaneous optical/X-ray observation ($\sim 100$ s) available to \citet{motch83} was too short to reveal any positive CCF components present.

\section{Discussion}

The main question that we wish to address is the origin of the rapidly variable optical power and the complex cross-correlation. Published models for the \xtej1118\ CCF include, among others: \citet{merloni00}, who invoke a magnetically-dominated corona, \citeauthor{esin01} (\citeyear{esin01}, a dominant advection-dominated flow [ADAF] with additional synchrotron), \citet[][ a pure jet]{markoff01}, \citet[][ feedback in a common jet+corona reservoir]{malzac04} and \citet[][ an ADAF and a jet dominating at different energies]{yuan05}. Most can explain the averaged broad-band energetics, and provide a qualitative description of the expected variability. But all agree that the origin and the details of the rapid variability patterns are likely to be complicated. The case of \gx339\ may well be similar, and a full investigation of the parameter space of the various models is beyond the scope of this Letter. Nevertheless, several important conclusions can be drawn from our simultaneous multi-wavelength data. 

\subsection{Reprocessing?}

\begin{figure}
  \begin{center}
    \includegraphics[angle=90,width=8.5cm]{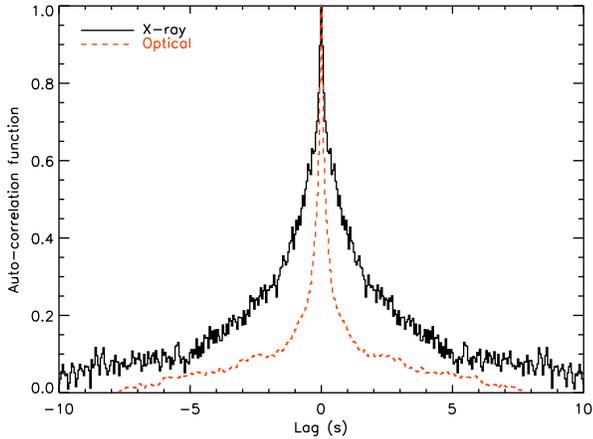}
\caption{The X-ray (black) and optical (red dashed) auto correlation functions (ACF) computed from the highest time resolution (50 ms) light curves on Night 1. The X-ray ACF is for the full-band PCA, and the optical refers to the $r'$ filter. Both are corrected for Poisson noise (dominant in X-rays).
 \label{fig:autocorr}}
  \end{center}
\end{figure}

The peak of the CCF time delay (150~ms) corresponds to a light-travel distance of $5000$ \gravrad\ [$\equiv GM/c^2$] for $M$$\gtsim$6~\Msun, too small for reprocessing on the companion star (\gx339\ has a binary separation of $\approx 25$ light-seconds). An auto-correlation analysis of the individual light curves can also be used to constrain the emission processes. Fig.~\ref{fig:autocorr} shows the computed X-ray and optical auto-correlation functions (ACFs). Low count-rate Poisson noise dominating the X-ray ACF at zero lag has been corrected-for by subtracting white noise from an X-ray power spectrum, followed by an inverse Fourier transform. The final X-ray ACF is broader than the optical one, similar to the result found in the case of \xtej1118. This also argues against a reprocessing origin (on the outer parts of the accretion disk, say) for the rapidly-variable component, at least as described by a simple, linear transfer function \citep[cf.][]{kanbach01}. 

\subsection{Behaviour around flares and dips}

Fig.~\ref{fig:likemalzac} shows that the behaviour of low- and high-energy X-ray (in this case, 2--5 keV and 5--20 keV) photon intensities is similar, when selected around full band PCA peaks and dips, with no obvious lag. We note that CCFs of light curves extracted in these energy ranges, with respect to the optical, showed little difference to the full-band result of Fig.~\ref{fig:crosscorr}. The optical light curves follow the CCF shape, as already mentioned. No significant colour ($g'-r'$) changes around the positions of X-ray flares/dips were detected.

\begin{figure}
  \begin{center}
    \includegraphics[angle=90,width=8.5cm]{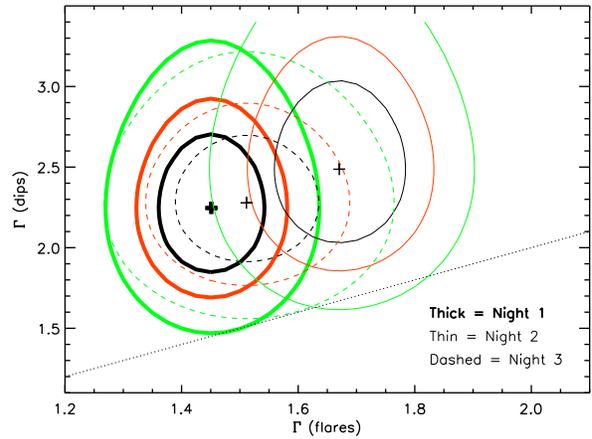}
\caption{
Photon-index ($\Gamma$) contours for single power-law fits to the extracted 3--30 keV flare and dip background-subtracted spectra, with absorption fixed to Galactic. The total resultant time interval for spectral extraction is $\approx 30$ s long for both flares and dips, and the net count rate is $\approx 40$ and 5 ct s$^{-1}$ respectively. Contour levels correspond to 68.3, 95 and 99\% for two interesting parameters, and include 1 per cent systematic errors. Results are shown for all three nights. The dotted diagonal line is the 1:1 line. 
 \label{fig:flares_dips_spectra}}
  \end{center}
\end{figure}

What about the source spectral behaviour? Using standard HEADAS tools, we extracted average X-ray spectra (and background) within short time-bins, $\sim \pm 50-150$ ms, centred on X-ray PCA full-band flares and dips (HEXTE was not used for this analysis, as the source is always background-dominated above $\sim 20$ keV). What we find is that the source hardens when it flares, and vice-versa. This is illustrated in Fig.~\ref{fig:flares_dips_spectra}, which shows the contours of independent fits to the extracted flares and dips spectra. A simple power-law (with absorption fixed to Galactic) was used to parametrize the spectral change between the two cases. The photon-index fitted to the flares spectrum is harder than that fitted to dips at 99 per cent confidence on all nights. Simulations were used to confirm that the results are robust to fluctuations in the background, which dominates above $\sim 5$ keV in the lower-flux dips spectrum.

On the other hand, no significant changes were detected in X-ray spectra extracted at the positions of {\em optical} peaks as compared to optical dips -- the spectral slopes were close to the slope for the average spectrum of the full dataset ($\Gamma\approx 1.6$). Extracting spectra 150 ms {\em before} optical flares and dips (as suggested by the CCF delay) showed a small but clear difference in the intensities of the X-ray flare and dip spectra, but no obvious change in spectral slope within the errors. This seems consistent with the low CCF peak strength -- i.e. every optical flare need not have been preceded by a locally-maximum X-ray flare, in spite of the average correlation.

\subsection{Implications}

The above analysis shows that the source is harder during X-ray flares, and vice-versa. In the context of a hot accretion disk corona model, this is consistent with increased Compton up-scattering when the source is (momentarily) brighter, and vice-versa. Fast Compton cooling timescales of tens of ms or smaller have been inferred in Galactic black holes in the low/hard state \citep[cf. ][]{guilbert82_nature}, for coronal electron temperatures $kT_e \sim 100$ keV and seed-photon disk-blackbody X-ray luminosities $< 0.01 L/L_{\rm Edd}$. Very similar physical parameters are inferred for \gx339\ during previous low/hard states \citep{joinet07, miller06}, and also contemporaneous with our observations (cf. \citealt{tomsick08}, who favour an inner disk radius [\rin] of $\sim 10$~\gravrad). Thus, similarly-fast cooling timescales are likely to hold for the case of \gx339, resulting in apparently-simultaneous flaring and hardening, which is what we observe.

As for the optical emission, we have already excluded simple reprocessing models. Bremsstrahlung can also be ruled out, as this would require a corresponding X-ray flux higher than that observed by several orders of magnitude. The most likely remaining physical mechanism is then synchrotron emission. Cyclo-synchrotron models in which several magnetized active regions with sizes of a few Schwarzschild radii contribute significantly to the dereddened optical flux have been investigated by \citet[ in addition cf. \citealt{fabian82}]{dimatteo99}. The active regions are characterized by $B\sim$ few $\times 10^6$ G, coronal optical depths $\tau\sim 0.2-1$ and $kT_e\sim 150-200$~keV. \citet{wardzinski00} have also discussed important modifications to such models. 

Within the context of these models, what could the shape and the time delay of the CCF correspond do? The steep decline that follows the CCF peak (Fig.~\ref{fig:crosscorr}) suggests the presence of some mechanism that cuts off the optical flares suddenly -- like infall of synchrotron-emitting blobs into the BH, say. A potential difficulty of this scenario is that a free-fall time of 150 ms for a 6--10 \Msun\ BH corresponds to a physical radius of $\sim 200-250$ \gravrad, which is larger than the value of \rin\ inferred by \citet{tomsick08}. Unless the disk has receded further or an atypically large corona is present, this does not correspond to an obvious physically meaningful scale.

An alternate hypothesis is that the rapid variability originates in non-thermal emission within a relativistic outflow or a jet. \citet{malzac04} invoked a magnetic energy reservoir that feeds both a jet component (dominating the optical), as well as an electron corona (dominating in X-rays) to explain the CCF of \xtej1118. Energy injection was modeled as shot flares. Feedback between the two components resulted in complex correlations, with the optical power being proportional to the differential of the X-ray flux. In such a model, the CCF shape and time lag are determined by the dissipation timescale of the process that injects energy into the jet. This should be largely independent of the exact injection mechanism, though the authors discuss the context of a magnetic energy reservoir. For our observations, a simple exponential ($\propto e^{-t/\tau}$) fit to the innermost part of the X-ray ACF (lags $<$0.5 s) in Fig.~\ref{fig:autocorr} gives a dissipation timescale $\tau=0.2\pm 0.05$ s (90\% error). This agrees with the observed optical delay, and suggests that such a differential correlation may apply to \gx339 as well.

If X-ray flares above the accretion disk trigger the dissipation and large-scale re-ordering of the poloidal magnetic field threading a jet, it is possible that the synchrotron (optical) emission will respond on timescales related to subsequent field build-up. Significant modulation of the poloidal field can occur on timescales orders of magnitude longer than the dynamical time of the inner accretion disk regions where this field originates \citep[][ see their Eq. 4]{livio03}. For $r_{\rm in} \sim 10$~\gravrad, the Keplerian dynamical time is 6 ms. Our observed delay of the optical (i.e., jet) component ($\sim 150$ ms) is 25 times longer, which can easily be accommodated within the picture of magnetic modulation. Synchrotron emission by plasma accelerated along the jet during this period, followed by rapid radiative cooling, may thus explain the positive CCF peak and delay. 

As for the anti-correlation troughs: if the initial X-ray flares that triggered magnetic re-ordering are related to field reconnection events in a coexistent corona, the coronal magnetic energy density will be released on the X-ray flaring timescale. This will lead to a decrease of any ambient synchrotron emission that is occurring within the corona itself, resulting in an anti-correlation of optical with X-rays. X-ray flaring is coherent over timescales of $\sim$ several seconds, as seen in the X-ray ACF (Fig.~\ref{fig:autocorr}), and this is also the total length of the anti-correlation seen in the CCF plot of Fig.~\ref{fig:crosscorr}. In short, interaction between the jet and coronal components may be the key to understanding the complex correlation structure. 

A testable prediction of this model is that the appearance and strength of the positive CCF signal should be intimately related to the prominence of the jet. Our observations probed the beginning of the low/hard state. As the sources enters deeper into this state and the jet establishes itself, its contribution to the overall energetics should grow (e.g. \citealt{fender04}), as should the related optical fractional variability rms. Support of this comes from the fact that we find an optical rms variability of $\sim 15$ per cent over the full light curve timescales, while \citet{motch83} found an rms of 50 per cent when the source was optically brighter (this is well-matched to the estimate of $\sim 50$ per cent by \citealt{corbel02}, based on broad-band photometry). Increased corona/jet coupling will result in a higher peak strength of the optical/X-ray CCF. Finally, if a stronger poloidal field is required to establish a stronger jet component, this may result in longer timescales for breaking and establishing this field component following reconnection flares. Changing CCF delays could then be used to directly probe the evolution of characteristic accretion/ejection structures in X-ray binaries. 

\section{Acknowledgements}

\ultracam\ is supported by STFC grant PP/D002370/1. PG acknowledges JSPS \& RIKEN Foreign Researcher Fellowships. He thanks R.P. Fender, A.A. Zdziarski \& T. Belloni for valuable comments, and the referee for a prompt report. TS and MD acknowledge Spanish Ministry grants AYA2004 02646 \& AYA2007 66887. 

\bibliographystyle{mnras}
\bibliography{gandhi_gx339_letter}

\label{lastpage}
\end{document}